\documentclass[a4paper,11pt]{article}
\usepackage{amsmath}
\usepackage{amsfonts}
\usepackage{amssymb, amsthm}
\numberwithin{equation}{section}

\begin{document}

\title{ Global existence and propagation speed for a generalized Camassa-Holm model with both dissipation and dispersion}

\author{Qiaoyi Hu$^{1,2}$\footnote{e-mail:
huqiaoyi@scau.edu.cn },~ Zhijun Qiao$^2$\footnote{e-mail: zhijun.qiao@utrgv.edu} \\$^1$Department of Mathematics, South China
Agricultural University,\\Guangzhou, Guangdong 510642, PR China\\
$^2$School of Mathematical and Statistical Science,\\ University of Texas -- Rio Grande Valley,\\ Edinburg, TX 78539, USA}

\date{}
\maketitle

\begin{abstract}
In this paper, we study a generalized Camassa-Holm (gCH) model with both dissipation and dispersion,% arbitrary dispersion coefficient,
which has (N + 1)-order nonlinearities and includes the following three integrable equations:
the Camassa-Holm, the Degasperis-Procesi, and the Novikov equations, as its reductions. We first present the local well-posedness and a
precise blow-up scenario of the Cauchy problem for the gCH equation. Then we provide several sufficient conditions that guarantee the global existence of the strong solutions to the gCH equation.
Finally, we investigate the propagation speed for the gCH equation when the initial data is compactly supported.\\


\end{abstract}

\section{Introduction}
\newtheorem{theorem1}{Theorem}[section]
\newtheorem{lemma1}{Lemma}[section]
\newtheorem{remark1}{Remark}[section]
\par
Recently, Himonas and Holliman \cite{HH1} studied  the following generalized Camassa-Holm equation
\begin{equation}\label{eq11}
u_{t}-u_{txx}-u^N u_{xxx}-\beta u^{N-1}u_x u_{xx}+(\beta+1)u^N u_x
=0
\end{equation}
where $N\in \mathbb{N},~\beta\in \mathbb{R}$, and proved
%was recently studied by Himonas et al.
 the local well-posedness and the nonuniform dependence of its Cauchy problem in Sobolev spaces $H^s$ with $s > \frac{3}{2}$.
  %were proved by Himonas and Holliman \cite{HH1}.
  Zhou and Mu studied % the local well-posedness,
  the persistence properties of strong solutions and the existence of its weak
solutions to Eq. (1.1) \cite{zhou2}.
Himonas and Thompson \cite{Him2} also showed the persistence properties and unique continuation of Eq. (1.1).  %were shown by Himonas and Thompson \cite{Him2}.
Eq. (1.1) is an evolution equation with $(N+1)-$order nonlinearities and includes %contains
three  remarkable integrable equations: the Camassa-Holm (CH) equation, the Degasperi-Procesi (DP) equation, and the Novikov equation (NE).

As $N = 1$ and $\beta = 2$, Eq.(1.1) reads as the well-known Camassa-Holm equation,
which models the unidirectional propagation of shallow water waves over a flat bottom,
% Here $u(t,x)$ stands for the fluid velocity at time $t$ in the spatial $x$ direction \cite{C-H,C-L,D-G-H,I-K, Iv,J}.
and is %It has a bi-Hamiltonian structure \cite{C1, F-F} and is
completely integrable with a bi-Hamiltonian structure \cite{C-H, F-F}. %Also there is a geometric
%interpretation of Eq.(1.1) in terms of geodesic flow on the
%diffeomorphism group of the circle \cite{C-K,Ko}.
The remarkable feature of the CH equation is its peaked soliton (peakon)
%waves are peaked with
solution in the form of $u(t,x) = ce^{-|x-ct|}$, where $c$ is a wave speed \cite{C-H}. %They are orbitally stable and
%interact like solitons \cite{B-S-S,C-S1}.
%The peaked traveling
%waves replicate a characteristic for the waves of great height --
%waves of largest amplitude that are exact solutions of the
%governing equations for water waves cf. \cite{Ci, C-Eb, T}.
The Cauchy problem of the CH equation has extensively been studied. For instance, its local
well-posedness  problem  for initial data $u_{0}\in
H^{s}$ with $s>\frac{3}{2}$ was studied in \cite{C4,C-E2,L-O, Rb}. More interestingly, the CH equation
has not only global strong solutions modelling permanent waves
\cite{Ca,C-E2} but also blow-up solutions modelling
wave breaking \cite{C2,C-E1, C-E2, C-E4,L-O,Rb}. On the
other hand, the CH equation has globally weak solutions with initial data $u_0\in
H^1$ \cite{B-C,C-M,X-Z} as well as the algebro-geometric solution \cite{Qiao-CMP}.

As $N = 1$ and $\beta = 3$,  Eq. (1.1) is cast to the Degasperi-Procesi (DP) equation \cite{DP}.
 %used an asymptotic integrability approach to isolate integrable third order equations, discovered the DP equation.
The DP equation is another integrable peakon model with quadratic nonlinearity, but with $3 \times 3 $ Lax pairs \cite{DHH}.
%shallow water dynamics. Degasperis, Holm and Hone \cite{DHH} prove the formal
%integrability of DP equation by constructing a Lax pair. They also show that it has a bi-Hamiltonian structure and an infinite sequence of conserved quantities, and admits exact peakon solutions.
%Also, like the CH equation, the DP equation possesses peakon solutions of the form $u(t,x) = ce^{-|x-ct|}$.
The well-posedness, global existence, and blow-up phenomena of DP equation may be seen in \cite{C-K,ELY,ELY1,HH2,HH2a,L-Z,Y1,Y2}, and the DP equation may also be generalized to an entire integrable hierarchy including both positive and negative flows \cite{Qiao-2004}.

As $N = 2$ and $\beta = 3$,  Eq. (1.1) is reduced to the Novikov equation (NE) \cite{Novikov},
which is also integrable with $3 \times 3 $ Lax pairs and with the peakon solution $u(t,x) = \sqrt{c}e^{-|x-ct|}$ as well \cite{HW1}.
%??? and also possesses many properties exhibited by Camassa-Holm
%type equations. One of which is the existence of peakon solutions of the form $u(t,x) = \sqrt{c}e^{-|x-ct|}$. In fact, Novikov produced about 20 integrable equations with quadratic nonlinearities
%that include the
%Camassa-Holm CH equation and the
%Degasperis-Procesi
%DP equation by using as test for integrability the existence of an infinite hierarchy of (quasi-) local higher symmetries.
In \cite{Novikov}, some integrable equations with cubic nonlinearities
were produced including the NE and the cubic CH or FORQ equation \cite{Fo,OR,Qiao-2006}.
The local well-posedness of the NE's Cauchy problem on both the line and the circle, and its global solution existence and global weak solutions were investigated in \cite{HH3,Ti,Wu,Wu2}.

\par In this paper, we study the Cauchy problem of the generalized Camassa-Holm equation (1.1) with both arbitrary dissipation and dispersion
%and compactly supported initial data $supp~u_0 \subset [a, b]$:
\begin{equation}\label{eq1}
\left\{\begin{array}{ll}u_{t}-u_{txx}-u^N u_{xxx}-\beta u^{N-1}u_x u_{xx}+(\beta+1)u^N u_x\\+k(1-\partial_{x}^{2})u_x+\lambda(1-\partial_{x}^{2})u
=0,\\&t > 0,\,x\in \mathbb{R}, \\
u(0,x) = u_{0}(x),&x\in \mathbb{R}, \\
\end{array}\right.
\end{equation}
where $supp~u_0 \subset [a, b]$ is a compactly supported initial data,
$k\in \mathbb{R}$ is an arbitrary dispersion coefficient, and $\lambda>0$ is a dissipative parameter.
A physical motivation of studying dissipative equations is to include energy dissipation mechanisms in the real world, which indeed occurs in our daily life.
 %that it is quite difficult to avoid energy dissipation mechanisms in a real world.
 In the literature there were many articles dealing with
nonlinear models with dissipation. Ott and Sudan \cite{O-S}
investigated how the KdV equation was modified by the presence of
dissipation and what the effect of such a dissipation was on soliton
solutions of the KdV equation. Ghidaglia \cite{G} studied the long time
behavior of solutions to the weakly dissipative KdV equation as a
finite dimensional dynamical system. Wu and Yin
discussed the blow-up, and blow-up rate and decay of solutions to
the weakly dissipative periodic CH equation (i.e. Eq.(1.2) with $N = 1, \beta = 2, k=0$) \cite{W-Y}. Thereafter, they also investigated
the blow-up and decay of solutions to the weakly
dissipative DP equation (i.e. Eq.(1.2) with $N = 1, \beta = 3, k=0$) on the line \cite{W-Y1}. Hu and Yin discussed
the blow-up and blow-up rate of solutions to a weakly dissipative periodic rod equation \cite{hu1}. Later, Hu studied the global existence and blow-up phenomena for a weakly dissipative periodic 2-component
Camassa-Holm system \cite{hu2}.
In 2014, Zhou, Mu and Wang \cite{zhou1} considered the weakly dissipative gCH equation
(i.e. Eq.(1.2) with $k=0$). However, the work done above was only involved in dissipative terms. Recently, Novruzov and Hagverdiyev \cite{Novru} analyzed the behavior of solutions to the dissipative CH equation with arbitrary dispersion coefficient (i.e. Eq.(1.2) with $N = 1, \beta = 2$).

In this paper, we discuss the local well-posedness, the global existence, and the propagation speed
of strong solutions to Eq. (1.2). Our study indicates that in comparison between Eq (1.1) ($k=\lambda=0$) and Eq. (1.2) ($k,\lambda\neq0$), %. Specifically,  we find that
some behaviors of
solutions to the gCH equation (1.2) with dissipation and dispersion  are similar
to the ones of Eq. (1.1), such as, the local
well-posedness and the blow-up scenario. However, the dissipative term $\lambda (u-u_{xx})$ and the dispersive term $k (u_x-u_{xxx})$  in Eq. (1.2) do have impacts on the global existence and the propagation speed of its solutions, which are shown below in Theorem 3.1 %. %, it is interesting to see that only the dissipative term affects the global existence of Eq. (1.2).
%Moreover, as shown in
and Theorem 4.1, respectively. In particular,
 %we see that
 the propagation speed
%of Eq. (1.2)
is seriously affected by both the dissipative
 parameter $\lambda$ and the dispersion coefficient $k$. On the other hand, it is worthy to note that
the main difficulty
in establishing the above results lies in controlling certain norms of $(N + 1)$-order nonlinearities.
%, and
%introducing an increasing diffeomorphism of $\mathbb{R}$.
%More specifically, the (N + 1)-order nonlinearities in Eq. (1.2) form the main obstacle to achieving the control.
In addition, some of our results cover the earlier corresponding results studied in \cite{Him2,Novru}.

\par
The paper is organized as follows. In the second section we
give some preliminary
%needed
results including the local well-posedness of Eq.(1.2), the precise
blowup scenario and some useful lemmas to study the global existence and
the propagation speed. In the third section we provide several global existence results, which reveal that Eq.(1.2) has
global solutions modelling permanent waves. In the fourth section
we study the propagation speed of strong solutions to Eq.(1.2) under the condition that the initial data has compact support.

\bigskip
\par
\noindent \textit{Notation.} Throughout the paper,
$\ast$ is referred to the convolution.
%We use $\mathbb{S}= \mathbb{R}/\mathbb{Z}$
%to represent the circle of unit length.
The norm in the Lebesgue space $ L^p(\mathbb{R})$ is
denoted by $ \| \cdot \|_{L^p}, $ while $ \| \cdot \|_{H^s}, \, s
\in \mathbb{R}$, stands for the norm in the classical Sobolev spaces $
H^s(\mathbb{R}),$ where $ 1 \leq p \leq
\infty. $

\section{Preliminaries}
\newtheorem{theorem2}{Theorem}[section]
\newtheorem{lemma2}{Lemma}[section]
\newtheorem{remark2}{Remark}[section]
\par
%Let us first briefly give the needed results
In this section, we display some necessary results in order to reach our goal. Let us first
present the local well-posedness for the Cauchy problem of Eq.(1.2) in $H^{s}(\mathbb{R}), s >\frac{3}{2}$.
Hence, we rewrite Eq.(1.2) in the form
of a quasi-linear evolution equation of hyperbolic type. Letting
$y:=u-u_{xx}$ yields%, we get:
\begin{equation}
\left\{\begin{array}{ll} y_{t}+ y_{x}(u^N+k)+\frac{\beta}{N} y(u^N)_{x}+ \lambda y &= 0,\\&t > 0,\,x\in \mathbb{R},\\
y(0,x)=u_{0}(x)-\partial^{2}_{x}u_{0}(x),&x\in
\mathbb{R}. \end{array}\right.
\end{equation}
Note that  $G(x):=\frac{1}{2}e^{-|x|}$ is the
kernel of $(1-
\partial^{2}_{x})^{-1}$. Then $(1-
\partial^{2}_{x})^{-1}f = G*f $ for all $f \in L^{2}(\mathbb{R})$ and $G \ast
y=u $. Therefore, %By the identity,
Eq.(2.1) can be reformulated in the following form:%as follows:
\begin{equation}
\left\{\begin{array}{ll} u_{t}+ (u^N+k)u_{x}+\partial_{x}G\ast
h+G\ast g= 0,\\&t
> 0,\,x\in \mathbb{R},\\ u(0,x) = u_{0}(x),&x\in \mathbb{R}, \end{array}\right.
\end{equation}
or an equivalent form:
\begin{equation}
\left\{\begin{array}{ll} u_{t}+(u^N+k)u_{x}=-\partial_{x}(1-\partial^{2}_{x})^{-1}h-(1-\partial_{x}^{2})^{-1}g,& t>0,\, x\in \mathbb{R}, \\
u(0,x)=u_{0}(x),&x\in \mathbb{R},\end{array}\right.
\end{equation}
where $$h:=\frac{\beta}{N+1}u^{N+1}+\frac{3N-\beta}{2}u^{N-1}u^2_{x}-\lambda u_x,$$ and
$$g:=\frac{(N-1)(\beta-N)}{2}u^{N-2}u_x^3+\lambda u .$$

\par Note that, the local well-posedness of the Cauchy problem for Eq. (1.2) (or Eq.(2.3)) in $H^s(\mathbb{R}), s>\frac{3}{2}$ with $\lambda=k =0$ can be obtained by Kato's semigroup theorem \cite{K} (see, for example,  \cite{zhou2}), or by applying the contraction-mapping principle (see, \cite{HH1}). %It is easy to see that the same result holds for Eq.(1.2). The proof of this fact repeats almost word for word the corresponding proof for the case $\lambda=k =0$, that is why we omit the further details and present corresponding result directly. We now present the following local well-posedness result.

\begin{theorem2}\label{wellposed}
Given $u_{0} \in H^{s}(\mathbb{R}),\;s>\frac{3}{2}$, there exists
a maximal $T>0$, and a unique solution $u$ to
Eq.(1.2) (or Eq.(2.3)) such that
$$
u=u(.,u_{0})\in C([0,T);H^{s}(\mathbb{R}))\cap
C^{1}([0,T);H^{s-1}(\mathbb{R})).
$$
Moreover, the solution depends continuously on the initial data,
i.e. the mapping $u_{0} \rightarrow u(.,u_{0}):H^{s}(\mathbb{R})
\rightarrow C([0,T);H^{s}(\mathbb{R}))\cap
C^{1}([0,T);H^{s-1}(\mathbb{R}))$ is continuous. Furthermore, $T$
may be chosen independent of $s$ in the following sense. If $u\in
C([0,T);H^{s}(\mathbb{R}))\cap C^{1}([0,T);H^{s-1}(\mathbb{R}))$ is
a solution to Eq.(1.2), and if $u_{0} \in H^{s^{'}}(\mathbb{R})$ for
some $s^{'} \neq s, s^{'}
> \frac{3}{2}$, then
$$u \in C([0,T);H^{s^{'}}(\mathbb{R}))\cap
C^{1}([0,T);H^{s-1}(\mathbb{R})). $$ In particular, if $u_{0}
\in C^{\infty}(\mathbb{R}) = \bigcap\limits_{s \geq
0}H^{s}(\mathbb{R})$, then $u \in C([0,T); C^{\infty}(\mathbb{R}))$.
\end{theorem2}

Next, we address the global existence of Eq. (1.2). To see this, let us recall the following useful lemmas.

\begin{lemma2}\cite{K2}
If $r > 0$, then $H^r \cap L^{\infty}$ is an algebra. Moreover
$$\|fg\|_{H^r}\leq c(\| f \|_{L^{\infty}}\| g \|_{H^r} + \| f
\|_{H^r}\| g \|_{L^{\infty}}),$$ where c is a constant depending
only on $r$.
\end{lemma2}

\begin{lemma2}\cite{K2}
If $r > 0$, then $$\|[\Lambda^r, f]g\|_{L^2}\leq c(\| f_x
\|_{L^{\infty}}\| \Lambda^{r-1}g \|_{L^2} + \| \Lambda^rf \|_{L^2}\|
g \|_{L^{\infty}}),$$ where c is a constant depending only on $r$.
\end{lemma2}

\begin{lemma2}\cite{C-M}
Let $g \in C^{m+2}(\mathbb{R}) $ and $g(0) = 0.$ Then for every
$\frac{1}{2} < r\leq m,$ we have
$$\|g(u)\|_r \leq \tilde{g}(\|u\|_{L^{\infty}})\|u\|_r,~~u\in H^r.$$
where $\tilde{g}$ is a monotone increasing function depending only on the $g$ and $r$.
\end{lemma2}

\begin{lemma2} \cite{K2} Let $f\in H^s, s>\frac{3}{2}.$ Then
$$ \|\Lambda^{-r}[\Lambda^{r+t+1},
M_f]\Lambda^{-t}\|_{L^2}\leq
c\|f\|_{H^s},\ \ \ \ \ |r|,\ |t|\leq s-1,$$ where
$M_f$ is the operator of multiplication by $f$ and $c$ is a
positive constant depending only on $r,t$.
\end{lemma2}

\begin{lemma2} \cite{K} Let r,t be real
numbers such that $-r<t\leq r$. Then
$$
\|fg \|_{H^{t}}\leq
c\|f\|_{H^{r}}\|g\|_{H^{t}},  \ \ \ \ if \
r>\frac{1}{2},\\
$$
$$
\|fg \|_{H^{t+r-\frac{1}{2}}}\leq
c\|f\|_{H^{r}}\|g\|_{H^{t}}, \ if \
r<\frac{1}{2},
$$
where c is a positive  constant depending on r, t.
\end{lemma2}

Let us now give the global existence result.
\begin{theorem2}
Let $u_{0}\in H^{s}$, $s > \frac{3}{2}$, and let $T$ be the maximal existence time of
the solution $u$ to Eq.(1.2)(or Eq. (2.3))
with the initial data $u_{0}$. If there exists $M >0$ such that
$$
\|u(t,\cdot)\|_{L^{\infty}}+\|u_{x}(t,\cdot)\|_{L^{\infty}}\leq
M,~~t\in[0,T),
$$
then the $H^s-$ norm of the solution $u$ does
not blow up in finite time.
\end{theorem2}
\begin{proof}
Let $u$ be the solution to Eq. (2.3) with the initial data $u_0 \in
H^s, s>\frac {3} {2},$  and let $T$ be the maximal
existence time of the corresponding solution $u$, which is
guaranteed by Theorem 2.1.

%Throughout this proof, $c > 0$ stands for
%a generic constant depending on $s, N, \beta$ and $\lambda$, and $C > 0$ stands for
%a generic constant depending on $c$ and $M$.

Applying the operator $\Lambda ^s$ to Eq. (2.3),
multiplying by $\Lambda ^s u$, and integrating over $\mathbb{R}$, we
obtain
\begin{align*}
&\frac{d}{dt}\|u\|_{s}^{2}=-2((u^N+k) u_{x},u)_{s}+2(u,f_{11})_{s}+2(u,f_{12})_{s}\\
&\phantom{\frac{d}{dt}\|u\|_{s}^{2}}:=I+II+III,
\end{align*}
where$$
f_{11}=-\Lambda^{-2}g,~~~~~~
f_{12}=-\partial_{x}\Lambda^{-2}h.
$$
Let us estimate $I,II$ and $III$. Noting $(\Lambda^{s}u_{x},\Lambda^{s}u)_{L^2}=0,$ we have
\begin{align*}
&|I|=|-2((u^N+k)u_{x},u)_{s}|=2|(\Lambda^{s}(u^Nu_{x}),\Lambda^{s}u)_{L^2}|\\
&\phantom{|I|}=2|([\Lambda^{s},u^N]\partial_{x}u,\Lambda^{s}u)_{L^2}+(u^N\Lambda^{s}\partial_{x}u,\Lambda^{s}u)_{L^2}|\\
&\phantom{|I|}\leq 2\|[\Lambda^{s},u^N]\partial_{x}u\|_{L^{2}}\|\Lambda^{s}u\|_{L^{2}}+|((u^N)_x\Lambda^{s}u,\Lambda^{s}u)_{L^2}|\\
&\phantom{|I|}\leq c(\|(u^N)_x\|_{L^{\infty}}\|\|\Lambda^{s-1}u_x\|_{L^{2}}+\|\Lambda^{s}u^N\|_{L^{2}}
\| u_{x}\|_{L^{\infty}})\|\Lambda^{s}u\|_{L^{2}}\\
&\phantom{|I|\leq}
+\|(u^N)_x\|_{L^{\infty}}|(\Lambda^{s}u,\Lambda^{s}u)_{L^2}|\\
&\phantom{|I|}\leq c(\|Nu^{N-1}u_x\|_{L^{\infty}}\|u\|_{H^s}+\tilde{g_1}(\|u\|_{L^{\infty}})\|u\|_{H^s}\|u_x\|_{L^{\infty}})
\|u\|_{H^s}+\|(u^N)_x\|_{L^{\infty}}|\|u\|_{H^s}^2\\
&\phantom{|I|}\leq
c(2NM^N+M\tilde{g_1}(M))\|u\|^{2}_{H^s}:=C\|u\|^{2}_{H^s},
\end{align*}
where we use Lemma 2.2 with $r=s,f=u^N,g=u_x$ when estimating $\|[\Lambda^{s},u^N]\partial_{x}u\|_{L^{2}}$ and Lemma 2.3 with $r=s,g_1(u)=u^N$ when estimating $\|\Lambda^{s}u^N\|_{L^{2}}$.

Obviously, when $N=1$, $|II|$ is easy to get. When $N=2$, we refer to the proof of Theorem 3.1 in \cite{Wu}. Without loss of generality, we assume that $N\geq3$.
\begin{align*}
&|II|=|2(f_{11}(u),u)_{s}|=2|(\Lambda^{s-2}(\frac{(N-1)(\beta-N)}{2}u^{N-2}u_x^3+\lambda u),\Lambda^{s}u)_{L^2}|\\
&\phantom{|II|}\leq2|([\Lambda^{s-1}(\frac{(N-1)(\beta-N)}{2}u^{N-2}u_x^3+\lambda u),\Lambda^{s-1}u)_{L^2}|\\
&\phantom{|II|}\leq c(|([\Lambda^{s-1},u^{N-2}]u_{x}^3,\Lambda^{s-1}u)_{L^2}+(u^{N-2}\Lambda^{s-1}u_{x}^3,\Lambda^{s-1}u)_{L^2}|+\|\Lambda^{s-1}u\|_{L^2}^2)\\
&\phantom{|II|}\leq c(\|[\Lambda^{s-1},u^{N-2}]u_{x}^3\|_{L^{2}}\|\Lambda^{s-1}u\|_{L^{2}}+\|u^{N-2}\Lambda^{s-1}u_{x}^3\|_{L^2}\|\Lambda^{s-1}u\|_{L^{2}}+\|u\|_{H^{s-1}}^2)\\
&\phantom{|II|}\leq
c(\|(u^{N-2})_{x}\|_{L^{\infty}}\|u_{x}^3\|_{{H^{s-1}}}+\|u_{x}^3\|_{L^{\infty}}\|u^{N-2}\|_{H^{s-1}})\|u\|_{H^{s-1}}
\\&\phantom{|II|\leq}+c(\|u^{N-2}\|_{L^{\infty}}\|u_x^3\|_{H^{s-1}}\|u\|_{H^{s-1}}+\|u\|_{H^{s-1}}^2)
\\
&\phantom{|II|}\leq
c[\|(N-2)u^{N-3}u_{x}\|_{L^{\infty}}\tilde{g_2}(\|u_x\|_{L^{\infty}})\|u_x\|_{H^{s-1}}
+\|u_x\|_{L^{\infty}}^3\|\tilde{g_3}(\|u\|_{L^{\infty}})\|u\|_{H^{s-1}}]\|u\|_{H^{s-1}}
\\&\phantom{|II|\leq}+c[\|u^{N-2}\|_{L^{\infty}}\tilde{g_2}(\|u_x\|_{L^{\infty}})\|u_x\|_{H^{s-1}}\|u\|_{H^{s-1}}+
\|u\|^{2}_{H^{s-1}}]\\
&\phantom{|II|}\leq c((N-2)M^{N-2}\tilde{g_2}(M)+M^3\tilde{g_3}(M)+M^{N-2}\tilde{g_2}(M)+1)\|u\|^{2}_{H^s}\\
&\phantom{|II|}:=C\|u\|^{2}_{H^s},
\end{align*}
where we use Lemma 2.2 with $r=s-1,f=u^{N-2},g=u_x^3$ when estimating $\|[\Lambda^{s-1},u^{N-2}]u_{x}^3\|_{L^{2}}$ and  Lemma 2.3 with $r=s-1, g_2(u_x)=u_x^3$ and $g_3(u)=u^{N-2}$ when estimating $\|u_x^3\|_{H^{s-1}}$ and $\|u^{N-2}\|_{H^{s-1}}$.

Note that $H^s$ and $H^{s-1}$ are algebraic with $s>\frac{3}{2}$. Hence
\begin{align*}
&|III|=|2(f_{12}(u),u)_{s}|\leq2\|f_{12}(u)\|_{H^s}\|u\|_{H^s}\\
&\phantom{|III|}\leq\|\frac{\beta}{N+1}u^{N+1}+\frac{3N-\beta}{2}u^{N-1}u^2_{x}-\lambda u_x\|_{H^{s-1}}\|u\|_{H^s}\\
&\phantom{|III|}\leq c\|u\|_{H^s}(\|u^{N+1}\|_{H^{s-1}}+\|u^{N-1}u_{x}^{2}\|_{H^{s-1}}+\|u_x\|_{H^{s-1}})\\
&\phantom{|III|}\leq c\|u\|_{H^s}(\tilde{g_4}(\|u\|_{L^{\infty}})\|u\|_{H^{s-1}}+\|u_{x}^2\|_{L^{\infty}}\|u^{N-1}\|_{H^s}
+\|u_{x}^2\|_{H^{s-1}}\|u^{N-1}\|_{L^{\infty}}+\|u\|_{H^s})\\
&\phantom{|III|}\leq c\|u\|_{s}[\tilde{g_4}(\|u\|_{L^{\infty}})\|u\|_{H^{s-1}}
+\|u_{x}^2\|_{L^{\infty}}\tilde{g_5}(\|u\|_{L^{\infty}})\|u\|_{H^s}\\
&\phantom{|III|\leq}
+ \tilde{g_6}(\|u_{x}\|_{L^{\infty}})\|u_x\|_{H^{s-1}}\|u^{N-1}\|_{L^{\infty}}+\|u\|_{H^s}]\\
&\phantom{|III|}\leq c(\tilde{g_4}(M)+M^2\tilde{g_5}(M)+M^{N-1}\tilde{g_6}(M))\|u\|^{2}_{H^s}\\
 &\phantom{|III|}:=
C\|u\|_{H^s}^{2},
\end{align*}
where we use Lemma 2.1 with $r=s-1$ when estimating $\|u^{N-1}u_{x}^{2}\|_{s-1}$ and  Lemma 2.3 with $r=s-1, g_4=u^{N+1}, g_5=u^{N-1}, g_6(u_x)=u_x^2, $ when estimating $\|u^{N+1}\|_{H^{s-1}}$, $\|u^{N-1}\|_{H^{s-1}}$ and $\|u_x^2\|_{H^{s-1}}$.

Therefore
\begin{align*}
&\frac{d}{dt}\|u\|_{H^s}^{2}\leq
C\|u\|_{H^s}^{2}.
\end{align*}
Application of the Gronwall's inequality and assumption of the
theorem lead to %yields
\begin{align*}
&\|u\|_{H^s}^{2}\leq
\exp(Ct)\|u_0\|_{H^s}^{2},
\end{align*}
which completes the proof of the theorem.
\end{proof}

Now, we come to present the precise blow-up scenario.
\begin{theorem2} \label{blowup}
If $u_{0}\in H^{s}(\mathbb{R})$, $s >
\frac{3}{2}$, then the solution $u = u(.,u_{0})$ of Eq.(1.2) blows up in
the finite time $T < +\infty$ if and only if
$$\limsup _{t \uparrow T} \|u_{x} (t,\cdot)\|_{L^{\infty}}
= + \infty ~ or \limsup _{t \uparrow T}\|u (t,\cdot)\|_{L^{\infty}}
= + \infty.
$$
\end{theorem2}
% and several
%sufficient conditions on the initial data that ensure strong
%solutions to Eq.(1.2) blow up in finite time.
\begin{proof}
Assume $u_{0}\in H^{s}$ for some $s\in
N,\;s\geq 2$. Multiplying both sides of Eq.(2.1) by $y$ and using integration by parts with respect to $x$, we obtain%get
\begin{eqnarray}\label{yL2}
\nonumber\int_{\mathbb{R}}2yy_t&=&-2\int_{\mathbb{R}}y_xu^Nydx
-\frac{2\beta}{N}\int_{\mathbb{R}}y^2(u^N)_xdx-2k\int_{\mathbb{R}}yy_xdx-2\lambda\int_{\mathbb{R}}y^2dx\\
\nonumber&=&(1-\frac{2\beta}{N})\int_{\mathbb{R}}y^2(u^N)_xdx-2\lambda\int_{\mathbb{R}}y^2dx\\
&=&(1-\frac{2\beta}{N})\int_{\mathbb{R}}y^2Nu^{N-1}u_xdx-2\lambda\int_{\mathbb{R}}y^2dx
\end{eqnarray}
which implies the following result:
%From the above equality, we see that
if $ \|u\|_{L^{\infty}}$ and $ \|u_{x}\|_{L^{\infty}}$ are bounded on $[0,T)$, i.e. there exists a
positive constant $M$ such that $\|u\|_{L^{\infty}},~~\|u_{x}\|_{L^{\infty}} \leq M$, then we have
\begin{align*}
&\frac{d}{dt}\int_{\mathbb{R}} y^{2}dx \leq
c(M^N+1)\int_{\mathbb{R}} y^2dx.
\end{align*}
Substituting $y=u-u_{xx}$ into the above inequality, noticing %leads to%By the definition of $y=u-u_{xx}$, we have
%$$
%\int_{\mathbb{R}}y^2 dx = \int_{\mathbb{R}}(u-u_{xx})^2dx=\int_{\mathbb{R}}(u^2+2u_x^2+u_{xx}^{2})dx.
%$$
%Hence, we obtain
$$
\|u\|_{H^2}^2 \leq \|y\|_{L^2}^2\leq 2\|u\|_{H^2}^2,
$$
and using %By the above inequality and by means of
the Gronwall's inequality, we arrive at%get
\begin{align*}
&\|u\|_{H^2}^2\leq \int_{\mathbb{R}}y^2dx \leq  e^{c(M^N+1)T}\int_{\mathbb{R}}y_0^2dx\\
&\phantom{\|u\|_{H^2}}\leq
2e^{c(M^N+1)T}\|u_0\|_{H^2}^2,
\end{align*}
which implies %The above inequality shows
that the $H^{2}$-norm of the solution to
Eq.(2.1) does not blow up in a finite time.
By the Sobolev's embedding theorem, Theorem 2.2 ensures that $\|u\|_{H^s}$ does not blow up in a finite time.

On the other hand, by Sobolev's imbedding theorem, if
$$\limsup _{t \uparrow T} \{\sup _{x \in \mathbb{R}}\|u_{x} (t,\cdot)\|_{L^{\infty}} \}
= + \infty ~ or \limsup _{t \uparrow T} \{\sup _{x \in \mathbb{R}}\|u (t,\cdot)\|_{L^{\infty}} \}
= + \infty,
$$
then the solution will blow up in a finite time.

 Applying Theorem 2.2 and a simple density argument, one may know that Theorem 2.3 holds for all
$s>\frac{3}{2}$.
\end{proof}

\begin{remark2}
Theorem 2.3 covers Lemma 5.1 in \cite{Him2}.
\end{remark2}

Let us now give the following key lemmas which will be used later on in some theorems.% sequel.
\begin{lemma2}\label{H1}
Let $u_{0} \in H^{1}({\mathbb{R}}),$ and $\beta=N+1$, then as long as the solution
$u(t)$ given by Theorem 2.1 exists for any $t \in [0,T)$, we have
$$
\| u\|^{2}_{H^1}=e^{-2\lambda t}\| u_{0}\|^{2}_{H^1}.
$$
\end{lemma2}
\begin{proof}
Multiplying both side of Eq.(2.1) by $u$ and integrating by parts,
we obtain
\begin{equation*}
\int_{\mathbb{R}} uy_{t}dx + \int_{\mathbb{R}}u u^{N}y_{x}dx
+k\int_{\mathbb{R}}uy_{x}dx+ \frac{N+1}{N} \int_{\mathbb{R}} u(u^N)_{x}ydx  + \int_{\mathbb{R}}\lambda yudx= 0.
\end{equation*}
Noticing %that
\begin{equation*}
\int_{\mathbb{R}} u^{N+1}y_{x}dx+\int_{\mathbb{R}}
\frac{N+1}{N} \int_{\mathbb{R}} u(u^N)_{x}ydx=0 \quad and \quad k \int_{\mathbb{R}}
uy_{x}dx=0,
\end{equation*} %Then,
we have
$$
\frac{1}{2}\frac{d}{dt}\int_{\mathbb{R}}(u^{2}+u^{2}_{x})dx+\lambda
\int_{\mathbb{R}}(u^{2}+u^{2}_{x})dx=0,
$$
namely,%or
$$
\frac{d}{dt}\| u\|^{2}_{H^1}+2\lambda \| u\|^{2}_{H^1} = 0,
$$
which implies the desired result in the lemma.
\end{proof}

%Letting $N=2\beta$ in Eq.(2.4), we get the following lemma.
%\begin{lemma2}\label{H2}
%Let $u_{0} \in H^{2}({\mathbb{R}}),$ and $N=2\beta$, then as long as the solution
%$u(t)$ given by Theorem 2.2 exists, for any $t \in [0,T)$, we have
%$$
%\| y\|_{L^2}^{2}=e^{-2\lambda t}\| y_{0}\|_{L^2}^{2}.
%$$
%\end{lemma2}

 So, given initial data $u_{0}\in H^{s}(\mathbb{R}),
$ $s\geq 2,$ Theorem \ref{wellposed} ensures the existence and uniqueness of
strong solutions to Eq.(\ref{eq1}). Let us consider the following initial value problem:
\begin{equation}\label{q}
\left\{\begin{array}{ll}q_{t} (t,x)= u^N(t,q(t,x))+k ,&t \in [0,T),\,x\in \mathbb{R}, \\
q(0,x) = x, &x\in \mathbb{R},\end{array}\right.
\end{equation}
where $u$ denotes the solution to
Eq.(\ref{eq1}) with the initial data $u_0$. Because $u(t,.)\in
H^2(\mathbb{S}) \subset C^m(\mathbb{S})$ with $0\leq m \leq
\frac{3}{2},$  $u \in C^{1}([0,T)\times
\mathbb{R},\mathbb{R}).$ Applying classical results in the
theory of ordinary differential equations may yield %, one can obtain
the following lemma, %results of $q$
which is a key in the proof of global
existence of solutions to Eq.(\ref{eq1}) in Theorem \ref{wellposed}.

\begin{lemma2}\cite{McKean} Let $u_{0}$ $\in H^{s}(\mathbb{R})
$, $s \geq 2,$ and let $T>0$ be the maximal existence time of
the solution $u $ to Eq.(\ref{eq1}) under the initial data $u_0$. Then Eq.(\ref{q}) has a unique solution $q \in
C^{1}([0,T)\times\mathbb{R},\mathbb{R}).$ Moreover, the map
$q(t,\cdot)$ is an increasing diffeomorphism of $\mathbb{R}$ with the following positive derivative with respect to $x$:
$$
q_{x}(t,x)=\exp\left(\int_{0}^{t}(Nu^{N-1}u_{x})(s,q(s,x))ds\right)>0,
~~~(t,x)\in[0,T)\times\mathbb{R}.
$$
\end{lemma2}
\begin{proof}
Differentiating Eq. (\ref{q}) with respect to $x$, we have
\begin{equation*}\label{qx}
\left\{\begin{array}{ll}\frac{d}{dt}q_{x} (t,x)=N u^{N-1}(t,q(t,x))q_x(t,x) ,&t \in [0,T),\,x\in \mathbb{R}, \\
q_x(0,x) = 1, &x\in \mathbb{R},\end{array}\right.
\end{equation*}
Solving the above equation leads to the desired result in the lemma.
\end{proof}

%It is easy to see from Eq. (2.1) that
\begin{lemma2}\label{y(t,q)}
Let $u_{0}$ $\in H^{s}(\mathbb{R})
$, $s \geq 2,$ and let $T>0$ be the maximal existence time of
 the solution $u$ to Eq.(\ref{eq1}) corresponding to the initial data $u_0$.
Then we have
\begin{eqnarray}
y(t,q(t,x))q_{x}^{\frac{\beta}{N}}(t,x) = y_{0}(x) e^{-\lambda t}, ~~~(t,x)\in
[0,T)\times\mathbb{R},
\end{eqnarray}
which implies%Moreover,
\begin{equation}
e^{-\lambda t}\|y_0\|_{L^{\frac{N}{\beta}}(\mathbb{R})}=\|y\|_{L^{\frac{N}{\beta}}(\mathbb{R})}. %\|y(t,q(t,\cdot))q_x^{\frac{\beta}{N}}(t,\cdot)\|_{L^2}\leq \|y(t,q(t,\cdot))\|_{L^2}\|q_x^{\frac{\beta}{N}}(t,\cdot)\|_{L^2}.
\end{equation}
In particular, if $N=2\beta$, we have%one can get
\begin{equation}\label{ylp}
e^{-\lambda t}\|y_0\|_{L^2(\mathbb{R})}=\|y\|_{L^2(\mathbb{R})}. %\|y(t,q(t,\cdot))q_x^{\frac{\beta}{N}}(t,\cdot)\|_{L^2}\leq \|y(t,q(t,\cdot))\|_{L^2}\|q_x^{\frac{\beta}{N}}(t,\cdot)\|_{L^2}.
\end{equation}
\end{lemma2}
%Motivated by McKean＊s deep observation for the Camassa每Holm equation in [19] we can do a similar %particle trajectory (see [22], for example) as
\begin{proof}
By Eq. (2.1), a direct calculation gives%and we have that
\begin{align*}
&~~\frac{d}{dt}[y(t,q(t,x))q_x^{\frac{\beta}{N}}]
=(y_t + y_x q_t)q_x^{\frac{\beta}{N}}
+\frac{\beta}{N}y q_x^{\frac{\beta}{N}-1}q_{tx}\\
&=[y_t + y_x (u^N+k)]q_x^{\frac{\beta}{N}}
+
\frac{\beta}{N}y q_x^{\frac{\beta}{N}-1}(u^N)_x q_x\\
&= (y_t + y_x u^N + k y_x+\frac{\beta}{N}y(u^N)_x)q_x^{\frac{\beta}{N}}
= -\lambda y q_x^{\frac{\beta}{N}}.\end{align*}
Solving for $y$ from the above equation, we obtain
$$
y(t,q(t,x))q_x^{\frac{\beta}{N}}(t,x)=y_0(x)e^{-\lambda t},
$$
which exactly yields Eq. (2.7): %By the above equality, we obtain
\begin{align*}\label{ylp}
&~~e^{-\lambda t}\|y_0\|_{L^{\frac{N}{\beta}}}^{\frac{N}{\beta}}= \|y(t,q(t,\cdot))q_x^{\frac{\beta}{N}}(t,\cdot)\|_{L^{\frac{N}{\beta}}}^{\frac{N}{\beta}}\\
%&=\int_{\mathbb{R}}|y(t,q(t,x))|^p (q_x(t,x)^{\frac{\beta}{N}})^pdx\\
&=\int_{\mathbb{R}}|y(t,q(t,x))|^{\frac{N}{\beta}} q_x(t,x)dx\\
&=\int_{\mathbb{R}}|y(t,z)|^{\frac{N}{\beta}} dz~~(\text{by~~setting}~~ q(t,x)=z).
\end{align*}
%Thus we get (2.7).
Apparently, letting $N=2\beta$ leads to Eq. (2.8).
\end{proof}

%???

\section{Global existence}
\newtheorem{theorem3}{Theorem}[section]
\newtheorem{lemma3}{Lemma}[section]
\newtheorem {remark3}{Remark}[section]
\newtheorem{corollary3}{Corollary}[section]
\par
In this section, we provide two global existence results for strong
solutions to Eq. (1.2).

\begin{theorem3}
Assume $u_0\in H^s(\mathbb{R}), s\geq 3,$ and %is such that the associated potential
$y_0=u_0-u_{0,xx}$ satisfies $y_0(x)\neq 0$ a.e. $x\in \mathbb{R}$ and $\|y_0\|_{L^2}\leq (\frac{2^{N+1}\lambda}{|N-2\beta|})^{\frac{1}{N}}$, where $\beta\neq \frac{N}{2}$. Then the solution $u(t, x)$ to Eq. (\ref{eq1}) globally exists.
\end{theorem3}
\begin{proof}
%Using (\ref{yL2}), and m
Multiplying both sides of Eq. (\ref{yL2}) %the last equality
by $e^{2\lambda t}$ yields%, we have
\begin{align*}
\frac{d}{dt}(e^{2\lambda t}\int_{\mathbb{R}}y^2dx)=(1-\frac{2\beta}{N})e^{2\lambda t}\int_{\mathbb{R}}y^2Nu^{N-1}u_xdx.
\end{align*}
Employing $G=\frac{1}{2}e^{-|x|}$, $G\ast y=u$, and $G_x\ast y=u_x$, and using the Young's inequality, we have
\begin{equation}\label{u}
\|u\|_{L^{\infty}}\leq \|G\|_{L^2}\|y\|_{L^2}=\frac{1}{2}\|y\|_{L^2},
\end{equation}
and
\begin{equation}\label{ux}
\|u_x\|_{L^{\infty}}\leq \|G_x\|_{L^2}\|y\|_{L^2}=\frac{1}{2}\|y\|_{L^2},
\end{equation}
which implies %By the above two inequalities, we get
\begin{align*}
\frac{d}{dt}(e^{2\lambda t}\int_{\mathbb{R}}y^2dx)\leq|N-2\beta|e^{2\lambda t}(\frac{1}{2}\|y\|_{L^2})^N\int_{\mathbb{R}}y^2dx.
\end{align*}
Thus, we obtain
\begin{align}\label{aaa}
\frac{d}{dt}(e^{2\lambda t}\int_{\mathbb{R}}y^2dx)\leq|N-2\beta|e^{2\lambda t}(\frac{1}{2}\|y\|_{L^2})^N\|y\|_{L^2}^{2} \nonumber\\
= \frac{|N-2\beta|}{2^N}e^{-N\lambda t}(e^{2\lambda t}\int_{\mathbb{R}}y^2dx)^{\frac{N+2}{2}}.
\end{align}
Let $f(t):=e^{2\lambda t}\int_{\mathbb{R}}y^2dx$. We claim that if $y_0(x)\neq 0$ a.e. $x\in \mathbb{R}$, then $f(t)>0$ for all $t\in [0,T).$
Apparently from Eq. (2.6), if $y_0(x)\neq 0$ a.e. $x\in \mathbb{R}$, then $y(t,q(t,x))\neq0$ a.e. $x\in \mathbb{R}$.
%By (2.6), we have
%\begin{equation*}
%e^{-\lambda t}\|y_0\|_{L^2}=\|y(t,q(t,\cdot))q_x^{\frac{\beta}{N}}(t,\cdot)\|_{L^2}>0.
%\end{equation*}
Therefore, we have
$$\int_{\mathbb{R}}y^2(t,z)dz = \int_{\mathbb{R}}y^2(t,q(t,x))q_x(t,x)dx >0,$$
which implies $f(t)>0$.
Solving (\ref{aaa}) for $f(t)$ leads to%, we have
\begin{align*}
\frac{d}{dt}(f(t)^{-\frac{N}{2}})\geq -\frac{N}{2}\frac{|N-2\beta|}{2^N}e^{-N\lambda t},
\end{align*}
which can be integrated with respect to $t$ to give%yields
\begin{eqnarray*}
f(t)^{-\frac{N}{2}}-f(0)^{-\frac{N}{2}}\geq -\frac{N}{2}\frac{|N-2\beta|}{2^N}\int_0^te^{-N\lambda s}ds\\
=\frac{N|N-2\beta|}{2^{N+1}}\frac{e^{-N\lambda  t}-1}{N\lambda},
\end{eqnarray*}
that is,
\begin{eqnarray*}
f(t)^{-\frac{N}{2}}\geq f(0)^{-\frac{N}{2}}-\frac{|N-2\beta|}{2^{N+1}\lambda}.
\end{eqnarray*}
Due to $\beta\neq \frac{N}{2}$ and $y_0(x)\neq 0$ a.e. $x\in \mathbb{R}$, therefore if $f(0)^{-\frac{N}{2}}-\frac{|N-2\beta|}{2^{N+1}\lambda}>0$, then we have $\|y_0\|_{L^2}\leq (\frac{2^{N+1}\lambda}{|N-2\beta|})^{\frac{1}{N}}$ and %then %we have
\begin{eqnarray*}
[f(0)^{-\frac{N}{2}}-\frac{|N-2\beta|}{2^{N+1}\lambda}]^{-1}\geq f(t)^{\frac{N}{2}},
\end{eqnarray*}
i.e.
\begin{eqnarray*}
[(\int_{\mathbb{R}}y_0^2dx)^{-\frac{N}{2}}-\frac{|N-2\beta|}{2^{N+1}\lambda}]^{-1}\geq \left(e^{2\lambda t}\int_{\mathbb{R}}y^2dx\right)^{\frac{N}{2}}.
\end{eqnarray*}
Therefore, we obtain
\begin{eqnarray}\label{yL2N}
\|y\|_{L^2}^N\leq(e^{-N\lambda t})[\|y_0\|_{L^2}^{-N}-\frac{|N-2\beta|}{2^{N+1}\lambda}]^{-1}.
\end{eqnarray}
By (\ref{u}), (\ref{ux}) and (\ref{yL2N}) %and the condition of the theorem $\|y_0\|_{L^2}\leq (\frac{2^{N+1}\lambda}{|N-2\beta|})^{\frac{1}{N}}$,
one may readily see that $u$ and $u_x$ are bounded:
\begin{eqnarray*}
\|u\|_{L^{\infty}},\|u_x\|_{L^{\infty}}\leq\frac{1}{2}\|y\|_{L^2}\leq\frac{1}{2}(e^{-\lambda t})[\|y_0\|_{L^2}^{-N}-\frac{|N-2\beta|}{2^{N+1}\lambda}]^{-\frac{1}{N}},
\end{eqnarray*}
which guarantee the global existence of the solution $u(t, x)$ to Eq. (\ref{eq1}).
\end{proof}

\begin{remark3}
Theorem 3.1 includes the result of Theorem 1 in \cite{Novru}, since the CH equation is a special case of Eq. (1.1) with $N=1$ and $\beta=2$.
\end{remark3}

\begin{remark3}
Theorem 3.1 shows that the dispersion term $ky_x$ does not affect the global existence of strong solution to Eq. (1.2), but the dissipation term $\lambda y$ does.
\end{remark3}

Let us now discuss the special case either $\beta=\frac{N}{2}$ or $\beta = N+1$. To do so, we shall use
%let us global existence of
%strong solutions of Eq.(1.2) By introducing
a family of
diffeomorphisms on a line to study the global existence of
strong solutions to Eq.(1.2).

\begin{theorem3}
Assume $u_0\in H^s(\mathbb{R}), s\geq 3,$ is a given initial data such that the associated potential $y_0=u_0-u_{0,xx}$ does not change sign. Then when $\beta = N+1$ or $\beta=\frac{N}{2}$ the corresponding solution $u(t, x)$ to Eq. (\ref{eq1}) globally exists.
\end{theorem3}

\begin{proof}
Without loss of generality, let us assume $y_0 \geq 0$. Lemma \ref{y(t,q)} tells us that $y(t,x) \geq 0$ for $x\in \mathbb{R}, t\geq0$ . Since $u = G\ast y$
and $G \geq 0$, one may immediately see %we have that
$u(t,x) \geq 0$ for all time $t \geq 0$.
%We now use the relation
Obviously with the aid of $u = \frac{1}{2} e^{-|x|} \ast y$, we may express $u$ and $u_x$ as follows:
\begin{align}\label{uux}
u(t,x)=\frac{e^{-x}}{2}\int_{-\infty}^x e^z y(t,z)dz +\frac{e^x}{2}\int_{x}^{\infty} e^{-z} y(t,z)dz,\\
u_x(t,x)=-\frac{e^{-x}}{2}\int_{-\infty}^x e^z y(t,z)dz +\frac{e^x}{2}\int_{x}^{\infty} e^{-z} y(t,z)dz.
\end{align}
%Assume that
Therefore, we have%Since $y_0$ is non-negative, then we obtain
\begin{align*}
u(t,x)+u_x(t,x)=e^x\int^{\infty}_x e^z y(t,z)dz\geq 0, \\
u(t,x)-u_x(t,x)=e^{-x}\int_{-\infty}^x e^z y(t,z)dz\geq 0.
\end{align*}
i.e. $|u_x|\leq u$.

When $\beta = N + 1$, Lemma \ref{H1} yields the following inequalities%, we have
\begin{equation}
|u_x|\leq u\leq \|u\|_{L^{\infty}}\leq \frac{1}{\sqrt{2}}\|u\|_{H^1} \leq \frac{1}{\sqrt{2}}\|u_0\|_{H^1}.
\end{equation}
When $N=2\beta$, Eq. (\ref{ylp}) leads to%, we get
\begin{equation}
|u_x|\leq u\leq \|u\|_{L^{\infty}}\leq \frac{1}{\sqrt{2}}\|u\|_{H^1} \leq \frac{1}{\sqrt{2}}\|u\|_{H^2}\leq \frac{1}{\sqrt{2}}\|y\|_{L^2}\leq \frac{1}{\sqrt{2}}\|y_0\|_{L^2},
\end{equation}
where we used the following relations% that
\begin{align*}
&~~\|u\|_{H^2}^2\leq\|y\|_{L^2}^2=\int_{\mathbb{R}}(u-u_{xx})^2dx =\int_{\mathbb{R}}(u^2-2uu_{xx}+u_{xx}^2)dx\\
&=\int_{\mathbb{R}}(u^2+2u_{x}^2+u_{xx}^2)dx\leq 2\|u\|_{H^2}^2.
\end{align*}
So, (3.8) and (3.9) combining with Theorem \ref{blowup} complete the proof.%desired result.
\end{proof}

\begin{remark3}
Theorem 3.2 covers Theorem 1.4 in \cite{Him2}.
\end{remark3}

\begin{remark3}
Comparing Theorem 3.1 with Theorem 3.2, we see a very interesting physical phenomenon. In the case of $\beta\neq \frac{N}{2}$, the size of the initial potential $y_0$ in $L^2$ space impacts the lifespan, namely, the small initial potential guarantees the global existence. However, in the case of $\beta = N+1$ or $\beta= \frac{N}{2}$, neither the
smoothness nor the size of the initial data affects the lifespan, but the sign of the initial potential $y_0$ does.
\end{remark3}

\section{Propagation speed}
\newtheorem{theorem4}{Theorem}[section]
\newtheorem{lemma4}{Lemma}[section]
\newtheorem {remark4}{Remark}[section]
\newtheorem{corollary4}{Corollary}[section]

In this section, we shall investigate impact of the dispersion coefficient $k$ and the dissipative parameter $\lambda$ on the propagation speed of solutions to Eq. (1.2).

\begin{theorem4}
 Assume that for some $T>0$ and $s\geq3$, $u \in C([0, T]; H^s(\mathbb{R}))$ is a strong solution of the initial value problem associated with Eq. (\ref{eq1}). If the initial data $u_0(x)$ is compactly supported on $[a,b]$, and if either $\beta=N$, where $N$ is a positive odd number or $N=1, 0\leq \beta \leq 3$, then for any $t\in[0, T]$, we have $u(t,x) =\frac{1}{2}E_+(t)e^{-x},$ for $x\geq q(t, b),$ and $u(x, t) =\frac{1}{2}E_-(t)e^{x},$ for $x\leq q(t, a),$ where $E_+(t)$ and $E_-(t)$ stand for continuous non-vanishing functions with $E_+(t) >0$ and $E_-(t)<0$ for $t\in[0, T]$.

\end{theorem4}
\begin{proof}
If $u_0$ is initially compact-supported on the closed interval $[a,b]$, then so is $y_0$. It follows from Lemma \ref{y(t,q)}
that $y( t,\cdot)$ is compactly supported within  the interval $[q(t,a), q(t,b)].$ %its support contained in the interval $[q(t,a), q(t,b)].$

Let us define the following two functions
\begin{align}
E_+(t)=\int_{q(t,a)}^{q(t,b)}e^z y(t,z)dz,~~E_-(t)=\int_{q(t,a)}^{q(t,b)}e^{-z} y(t,z)dz.
\end{align}
Then by (\ref{uux})-(3.7) %the definition of $E_+, E_-$ and the compactness of $y$,
we have %that
\begin{align}\label{uu}
\nonumber u(t,x)=\frac{e^{-x}}{2}E_+(t),~~x>q(t,b), \\
 u(t,x)=\frac{e^{x}}{2}E_-(t),~~x<q(t,a),
\end{align}
which generate the following derivative formulas%therefore from differentiating (\ref{uu}) directly we get
\begin{align}
\nonumber \frac{e^{-x}}{2}E_+(t)=u(t,x)=-u_x(t,x)=u_{xx}(t,x),~~x>q(t,b), \\
\nonumber \frac{e^{x}}{2}E_-(t)=u(t,x)=u_x(t,x)=u_{xx}(t,x),~~x<q(t,a).
\end{align}
Since $u(0,\cdot)$ is compactly supported on the interval $[a, b]$, this immediately gives us $E_+(0) = E_-(0) = 0$.

Noticing that $y(t,\cdot)$ is compactly supported on the interval $[q(t,a), q(t, b)]$, for each fixed $t$ we have
\begin{align}
 \frac{dE_+(t)}{dt}=\int_{q(t,a)}^{q(t,b)}e^z y_t(t,z)dz=\int_{-\infty}^{\infty}e^z y_t(t,z)dz,
\end{align}
which yields %Thus, we have
\begin{align*}
 &\frac{dE_+(t)}{dt}=\int_{q(t,a)}^{q(t,b)}e^z y_t(t,z)dz\\
 &=\int_{-\infty}^{\infty}e^z y_t(t,z)dz\\
 &=-\int_{-\infty}^{\infty}(-y_xu^N-\frac{\beta}{N}y(u^N)_x-ky_x-\lambda y)e^zdz\\
 &=\frac{\beta}{N+1}\int_{-\infty}^{\infty}e^zu^{N+1}dz
 +\frac{3N-\beta}{2}\int_{-\infty}^{\infty}e^zu^{N-1}u_z^2dz\\
& ~~+\frac{(N-\beta)(N-1)}{2}\int_{-\infty}^{\infty}e^zu^{N-2}u_z^3dz
-(\lambda-k)\int_{-\infty}^{\infty}e^zydz.
\end{align*}
Let \begin{align*}F:=\frac{\beta}{N+1}\int_{-\infty}^{\infty}e^zu^{N+1}dz
 +\frac{3N-\beta}{2}\int_{-\infty}^{\infty}e^zu^{N-1}u_z^2dz\\
 +\frac{(N-\beta)(N-1)}{2}\int_{-\infty}^{\infty}e^zu^{N-2}u_z^3dz.\end{align*}
 If either $\beta=N$ ($N$ is a positive odd number) or $N=1, 0\leq \beta \leq 3$, then
  $$
 \frac{d}{dt}E_+(t)+(\lambda-k)E(t)=F\geq 0,
 $$
 which implies
   $$
 \frac{d}{dt}(e^{(\lambda-k)t}E_+(t))\geq 0,
 $$
 that is
   $$
e^{(\lambda-k)t}E_+(t)\geq 0.
 $$
Therefore, $E_+(t)\geq 0.$

Adopting the similar procedure for $E_-(t)$ as above, we can arrive at %have
\begin{align*}
 &\frac{dE_-(t)}{dt}=\int_{q(t,a)}^{q(t,b)}e^{-z} y_t(t,z)dz=-F+(-\lambda-k)\int_{-\infty}^{\infty}e^{-z}ydz.
\end{align*}
So, $E_-(t)\leq 0.$
This completes the proof of Theorem 4.1.
\end{proof}

%\begin{remark4}
%When $N=1$ and $\beta=2$, Theorem 4.1 covers part of the results of Theorem 2 in \cite{Novru}.
%\end{remark4}

\begin{remark4}
From the proof of Theorem 4.1, we see that there are significant differences between the cases $k, \lambda\neq 0$ and $k,\lambda=0$. More precisely, the first order derivatives of the functions $E_+(t)$ and $E_-(t)$ in the case $k, \lambda \neq 0$, in general, can change their signs. While Theorem 4.1 in \cite{Him2} with $k=\lambda =0$ showed that $E_+(t)$ and $E_-(t)$ are strictly decreasing for $t \in [0, T)$.
\end{remark4}

\bigskip
\noindent\textbf{Acknowledgments} This work was supported by
NSFC (Nos. 11401223, 11171295, and 61328103), Guangdong NSF (No. 2015A030313424) and China Scholarship Council.
 The first author would like to thank Professor Zhijun Qiao for his kind hospitality and encouragement during her visit in the University of Texas - Rio Grande Valley, and the second author thanks the U.S. Department of Education GAANN project (P200A120256) for supporting the UTPA mathematics graduate program.

% The author thanks the referees for valuable
%comments and suggestions.


\begin{thebibliography}{99}

%\bibitem{B-S-S}
%R. Beals, D. Sattinger, and J. Szmigielski, Multipeakons and a
%theorem of Stieltjes, {\it Inverse Problems,} {\bf 15} (1999),
%1--4.

%\bibitem{B-B-M}
%T. B. Benjamin, J. L. Bona, and J. J. Mahony, Model equations for
%long waves in nonlinear dispersive systems, {\it Phil. Trans. R.
%Soc. (London),} {\bf 272} (1972), 47--78.

\bibitem{B-C}
A. Bressan and A. Constantin, Global conservative solutions of the
Camassa-Holm equation, {\it Arch. Rat. Mech. Anal.,} {\bf 183}
(2007), 215--239.

\bibitem{C-H}
R. Camassa and D. Holm, An integrable shallow water equation with
peaked solitons, {\it Phys. Rev. Letters,} {\bf 71} (1993),
1661--1664.

\bibitem{C-H-H}
R. Camassa, D. Holm and J. Hyman, A new integrable shallow water
equation, {\it Adv. Appl. Mech.,} {\bf 31} (1994), 1--33.

\bibitem{C-K}
G. Coclite and K. Karlsen, On the well-posedness of the Degasperis-Procesi
equation, {\it J. Funct. Anal.,} {\bf 233} (2006), 60--91.

\bibitem{C1}
A. Constantin, The Hamiltonian structure of the Camassa-Holm
equation, {\it Exposition. Math.,} {\bf 15} (1997), 53--85.

\bibitem{C4}
A. Constantin, The Cauchy problem for the periodic Camassa-Holm
equation, {\it J. Differential Equations,} {\bf 141} (1997),
218-235.

\bibitem{C2}
A. Constantin, On the blow-up of solutions of a periodic shallow
water equation, {\it J. Nonlinear Sci.,} {\bf 10} (2000), 391-399.

\bibitem{Ca}
A. Constantin, Existence of permanent and breaking waves for a
shallow water equation: a geometric approach, {\it Ann. Inst.
Fourier, Grenoble,} {\bf 50} (2000), 321--362.

\bibitem{C3}
A. Constantin, On the scattering problem for the Camassa-Holm
equation, {\it Proc. R. Soc. London A,} {\bf 457} (2001),
953--970.

\bibitem{Ci}
A. Constantin, The trajectories of particles in Stokes waves, {\it
Invent. Math.,} {\bf 166} (2006), 523--535.

\bibitem{C-E1}
A. Constantin and J. Escher, Wave breaking for nonlinear nonlocal
shallow water equations, {\it Acta Mathematica,} {\bf 181} (1998),
229--243.

\bibitem{C-E2}
A. Constantin and J. Escher, Well-posedness, global existence, and
blowup phenomena for a periodic quasi-linear hyperbolic equation,
{\it Comm. Pure Appl. Math.,} {\bf 51} (1998), 475--504.

%\bibitem{C-E3}
%A. Constantin and J. Escher, On the blow-up rate and the blow-up
%of breaking waves for a shallow water equation, {\it Math. Z.,}
%{\bf 233} (2000), 75--91.

\bibitem{C-E4}
A. Constantin and J. Escher, On the structure of a family of
quasilinear equations arising in a shallow water theory, {\it
Math. Ann.,} {\bf 312} (1998), 403--416.

\bibitem{C-Eb}
A. Constantin and J. Escher, Particle trajectories in solitary
water waves, {\it Bull. Amer. Math. Soc.,} {\bf 44} (2007),
423--431.

%\bibitem{C-K}
%A. Constantin and B. Kolev, Geodesic flow on the diffeomorphism
%group of the circle, {\it Comment. Math. Helv.,} {\bf 78} (2003),
%787--804.

%\bibitem{C-L}
%A. Constantin and D. Lannes, The hydrodynamical relevance of the
%Camassa-Holm and Degasperis-Procesi equations, {\it Arc. Rat.
%Mech. Anal.,} {\bf 192} (2009), 165--186.

%\bibitem{C-Mc}
%A. Constantin and H. P. McKean, A shallow water equation on the
%circle, {\it Comm. Pure Appl. Math.,} {\bf 52} (1999), 949--982.

%\bibitem{C-M}
%A. Constantin and L. Molinet, Global weak solutions for a shallow
%water equation, {\it Comm. Math. Phys.,} {\bf 211} (2000), 45--61.

%\bibitem{C-S1}
%A. Constantin and W. A. Strauss, Stability of peakons, {\it Comm.
%Pure Appl. Math.,} {\bf 53} (2000), 603--610.

\bibitem{C-M}
 A. Constantin and L. Molinet, The initial value problem for a generalized
Boussinesq equation, {\it Differential and Integral equations} {\bf 15} (2002) 1061--1072.

%\bibitem{C-S2}
%A. Constantin and W. A. Strauss, Stability of a class of solitary
%waves in compressible elastic rods, {\it Phys. Lett. A,} {\bf 270}
%(2000), 140--148.

\bibitem{DP}
A. Degasperis and M. Procesi, Asymptotic integrability symmetry and perturbation theory, (Rome,
1998), {\it World Sci. Publ.,} (1999), 23--37.

\bibitem{DHH}
A. Degasperis, D. Holm and A. Hone, A new integrable equation with
peakon solution, {\it Theoret. and Math. Phys,} {\bf 133} (2002) 1463--1474.


%\bibitem{D}
%H. H. Dai, Model equations for nonlinear dispersive waves in a
%compressible Mooney-Rivlin rod, {\it Acta Mech.,} {\bf 127}
%(1998), 193--207.

%\bibitem{D-H}
%H. H. Dai and Y. Huo, Solitary shock waves and other travelling
%waves in a general compressible hyperelastic rod, {\it Proc. R.
%Soc. London A,} {\bf 456} (2000), 331--363.

%\bibitem{D-E-G}
%R. K. Dodd, J. C. Eilbeck, J. D. Gibbon, and H. C. Morris, Solitons
%and Nonlinear Wave Equations, Academic Press, New York, 1984.

%\bibitem{D-G-H}
%H. R. Dullin, G. A. Gottwald, and D. D. Holm, An integrable
%shallow water equation with linear and nonlinear dispersion, {\it
%Phys. Rev. Letters,} {\bf 87} (2001), 4501--4504.

\bibitem{hu1}
Q. Hu and Z. Yin, Blowup and blowup rate of solutions to a weakly dissipative periodic rod equation,
{\it J.Math. Phys.,} {\bf 50} 083503 (2009), 1--16.

\bibitem{hu2}
Q. Hu, Global existence and blow-up phenomena for a weakly dissipative periodic 2-component Camassa-Holm system,
{\it J. Math. Phys.,} {\bf 52} 103701 (2011), 1--13.

%\bibitem{hu3}
%Q. Hu, Global existence and blow-up phenomena for a weakly dissipative 2-component Camassa-Holm system,
%{\it Appl. Anal.,} {\bf 92,} (2013), 398--410.

\bibitem{ELY}
J. Escher, Y. Liu and Z. Yin, Global weak solutions and blow-up structure
for the Degasperis-Procesi equation, {\it J. Funct. Anal.}, {\bf
241} (2006), 257--485.

\bibitem{ELY1}
J. Escher, Y. Liu and Z. Yin, Shock waves and blow-up phenomena for
the periodic Degasperis-Procesi equation, {\it Indiana Univ. Math. J.}, {\bf
56} (2007), 87--117.

\bibitem{Fo} A. Fokas, On a class of physically important integrable equations, {\it Physica D} {\bf 87} (1995), 145-150.

\bibitem{F-F}
A. Fokas and B. Fuchssteiner, Symplectic structures, their B{\"
a}cklund transformation and hereditary symmetries, {\it Physica
D,} {\bf 4} (1981), 47--66.

\bibitem{G}
J. Ghidaglia, Weakly damped forced Korteweg-de Vries equations
behave as a finite dimensional dynamical system in the long time,
{\it J. Differential Equations,} {\bf 74} (1988), 369--390.

%\bibitem{GY1}
%C. Guan, Z. Yin, Global weak solutions for a modified
%two-component Camassa每Holm equation, {\it Ann. I. H. Poincar谷 每 AN,} {\bf 28} (2011), 623--641.

%\bibitem{hu}
%Q. Hu and Z. Yin, Blowup phenomena for a new periodic nonlinearly
%dispersive wave equation, {\it Math. Nachr.,} {\bf In press}.

\bibitem{HH1}
 A. Himonas and C. Holliman, The Cauchy problem for a generalized Camassa-Holm equation,
{\it Advances in Differential Equations,} {\bf 19} (2013), 161--200.


\bibitem{HH2}
 A. Himonas and C. Holliman, On well-posedness of the Degasperis-Procesi equation, {\it Discrete
Contin. Dyn. Syst.,} {\bf 31} (2011), 469--488.

\bibitem{HH2a}
A. Himonas, C. Holliman and K. Grayshan, Norm inflation and ill-posedness for the Degasperis-Procesi equation, {\it Comm. Partial Differential Equations} {\bf 39} (2014) 2198--2215.

\bibitem{HH3}
A. Himonas and C. Holliman, The Cauchy problem for the Novikov equation, {\it Nonlinearity,} {\bf 25}
(2012), 449--479.

\bibitem{Him2}
A. Himonas and R. Thompson, Persistence properties and unique continuation
for a generzlized Camassa-Holm Equation, {\it J. Math. Phys.,} {\bf 55} (2014), 091503.

%\bibitem{I-K}
%D. Ionescu-Krus, Variational derivation of the Camassa-Holm
%shallow water equation, {\it J. Nonlinear Math. Phys.,} {\bf 14}
%(2007), 303--312.

%\bibitem{Iv}
%R. I. Ivanov, Water waves and integrability, {\it Philos. Trans.
%Roy. Soc. London A,} {\bf 365} (2007), 2267--2280.

%\bibitem{J}
%R. S. Johnson, Camassa-Holm, Korteweg-de Vries and related models
%for water waves, {\it J. Fluid. Mech.,} {\bf 457} (2002), 63--82.

%\bibitem{Ko}
%B. Kolev, Bi-Hamiltonian systems on the dual of the Lie algebra of
%vector fields of the circle and periodic shallow water equations,
%{\it Philos. Trans. Roy. Soc. London A}, {\bf 365} (2007),
%2333-2357.


\bibitem{HW1} A. Hone and J. Wang, Integrable peakon equations with cubic nonlinearity, {\it J. Phys. A: Math. Theor.} {\bf 41} (2008), 372002--372010.

\bibitem{K}
T. Kato, Quasi-linear equations of evolution, with applications to
partial differential equations, in: Spectral Theory and Differential
Equations, Lecture Notes in Math., Springer Verlag, Berlin, {\bf
448} (1975), 25--70.

\bibitem{K2}
T. Kato and G. Ponce, Commutator estimates and Navier-Stokes
equations, {\it Comm. Pure Appl. Math.,} {\bf 41} (1988), 203--208.

%\bibitem{Kato 1}
%T. Kato, Quasi-linear equations of evolution, with applications to
%partial differential equations, in "Spectral Theory and Differential
%Equations", Lecture Notes in Math., Vol. 448, Springer Verlag,
%Berlin, (1975), 25--70.

%\bibitem{Kato 2}
% T. Kato, On the Cauchy problem for the generalized Korteweg-de Vries equation,
%in; Studies in Applied Mathematics, in: Adv. Math. Suppl. Stu.,vol. 8,
%Academic Press, New York, 1983, pp. 93每128.

\bibitem{L-O}
Y. Li and P. Olver, Well-posedness and blow-up solutions for an
integrable nonlinearly dispersive model wave equation, {\it J.
Differential Equations,} {\bf 162} (2000), 27--63.

\bibitem{L-Z}
Y. Li and P. Olver, Well-posedness and blow-up solutions for phenomona for the
Degasperis-Procesi equation, {\it Comm. Math. Phys.,} {\bf 267} (2006), 801--820.

\bibitem{McKean}
H. McKean, Breakdown of a shallow water equation, {\it Asian J. Math.,} {\bf 2(4)} (1998) 867--874.

\bibitem{Novikov}
V. Novikov, Generalizations of the Camassa-Holm equation, {\it J. Phys. A,} {\bf 42} (2009), 342002, 14
pages.

\bibitem{Novru}
E. Novruzova and A. Hagverdiyevb, On the behavior of the solution of the dissipative Camassa-Holm equation with the arbitrary dispersion coefficient {\it J. Differential Equations,} {\bf 257} (2014), 4525--4541.


\bibitem{OR}
P. Olver and P. Rosenau, Tri-Hamiltonian duality between solitons and solitary-wave solutions having compact support, {\it Phys. Rev. E,} {\bf 53} (1996), 1900--1906.

\bibitem{O-S}
E. Ott and R. Sudan, Damping of solitary waves, {\it Phys.
Fluids,} {\bf 13} (1970), 1432--1434.


\bibitem{Qiao-CMP} Z. Qiao, The Camassa-Holm hierarchy, related $N$-dimensional integrable systems and algebro-geometric solution on a symplectic submanifold,
{\it Commun. Math. Phys.} {\bf 239} (2003) 309 -- 341.

\bibitem{Qiao-2004} Z. Qiao, Integrable hierarchy (the DP hierarchy), $3 \times 3 $ constrained systems, and parametric and stationary solutions, {\it Acta Applicandae Mathematicae} {\bf 83} (2004), 199-220.
%\bibitem{Pazy}
%A. Pazy, Semigroup of Linear Operators and Applications to Partial
%Differential Equations, Springer-Verlag, New York, 1983.

\bibitem{Qiao-2006} Z. Qiao, A new integrable equation with cuspons and W/M-shape-peaks solitons,  {\it J. Math. Phys.} {\bf 47} (2006), 112701 -- 112710.
    %New integrable hierarchy, its parametric solutions, cuspons, one-peak solutions, and M/W-shape peak solitons, {\it J. Math. Phys.} {\bf 48} (2007), 082701 -- 082711.
%\bibitem{Q11}Z.J. Qiao and X.Q. Li,  An integrable equation with nonsmooth solitons, Theor. Math. Phys. 167 584-589 (2011).


\bibitem{Rb}
G. Rodriguez-Blanco, On the Cauchy problem for the Camassa-Holm
equation, {\it Nonlinear Anal.,} {\bf 46} (2001), 309--327.

%\bibitem{S-S}
%P. Souganidis and W. A. Strauss, Instability of a class of
%dispersive solitary waves, {\it Proc. R. Soc. Edinburgh,} {\bf
%114A} (1990), 195--212.

\bibitem{T}
J. Toland, Stokes waves, {\it Topol. Methods Nonlinear Anal.,}
{\bf 7} (1996), 1--48.

\bibitem{Ti}
F. Ti$\breve{g}$lay, The periodic Cauchy problem for Novikov's
equation, {\it Int. Math. Res. Not.,} (2010) rnq267, 1--16.

\bibitem{W-Y}
S. Wu and Z. Yin, Blow-up, blow-up rate and decay of the solution
of the weakly dissipative Camassa-Holm equation, {\it J. Math.
Phys.,} {\bf 47}, 013504 (2006), 1--12.

\bibitem{W-Y1}
S. Wu and Z. Yin, Blow-up and decay of the solution of the weakly
dissipative Degasperis-Procesi equation, {\it SIAM J. Math.
Anal.,} {\bf 40} (2008), 475--490.

\bibitem{Wu}
X. Wu and Z. Yin, Well-posedness and global existence for the
Novikov equation, {\it Ann. Sc. Norm. Sup. Pisa CI. Sci.,} {\bf 5} (2012), 707--727.

\bibitem{Wu2}
X. Wu and Z. Yin, Global weak solutions for the Novikov equation,
{\it J. Phys. A: Math. Theor.,} {\bf 44} (2011), 055202 1--17.

\bibitem{X-Z}
Z. Xin and P. Zhang, On the weak solutions to a shallow water
equation, {\it Comm. Pure Appl. Math.,} {\bf 53} (2000),
1411--1433.

\bibitem{Y1}
Z. Yin, Global existence for a new periodic integrable equation, {\it J. Math. Anal. Appl.,} {\bf 283}
(2003), 129--139.

\bibitem{Y2}
Z. Yin, On the Cauchy problem for an integrable equation with peakon solutions, {\it Illinois J. Math.,} {\bf 47} (2003), 649--666.

\bibitem{zhou2}
S. Zhou and C. Mu, The properties of solutions for a generalized b-family
equation with peakons, {\it J Nonlinear Sci,} {\bf 23} (2013), 863--889.

\bibitem{zhou1}
S. Zhou, C. Mu and L. Wang, Well-posedness, blow-up phenomena and global existence for the generalized b-equation with higher-order nonlinearities and weak diddipation, {\it Discrete Contin. Dyn. Syst. Ser. A,} {\bf 34} (2014), 843--867.

\end{thebibliography}
\end{document}